\documentstyle[prl,aps,multicol,epsf]{revtex}
\renewcommand{\narrowtext}{\begin{multicols}{2} \global\columnwidth20.5pc}
\renewcommand{\widetext}{\end{multicols} \global\columnwidth42.5pc} 

\begin{document}

\newcommand{\be}{\begin{equation}}
\newcommand{\ee}{\end{equation}}
\newcommand{\bea}{\begin{eqnarray}}
\newcommand{\eea}{\end{eqnarray}}
\newcommand{\nt}{\narrowtext}
\newcommand{\wt}{\widetext}

\title{Elusive physical electron propagator in $QED$-like effective theories}

\author{D. V. Khveshchenko}

\address{Department of Physics and Astronomy, University of North
Carolina, Chapel Hill, NC 27599}
\maketitle

\begin{abstract}
We study the previously conjectured form of the physical electron propagator
and its allegedly Luttinger type of behavior
in the theory of the pseudogap phase of high-temperature copper-oxide superconductors
and other effective $QED$-like models.
We demonstrate that, among a whole family of seemingly gauge-invariant 
functions, the conjectured "stringy ansatz" for the electron propagator is the only 
one that is truly invariant. However, contrary to the results of the earlier works,
it appears to have a negative anomalous dimension, which makes it a rather poor candidate
to the role of the physical electron propagator.
Instead, we argue that the latter may, in fact, feature a "super-Luttinger" 
behavior characterized by a faster than any 
power-law decay $G_{phys}(x)\propto\exp(-const\ln^2|x|)$.
\end{abstract}

\section{Introduction}

In a generic $(1+1)$-dimensional (hereafter 2D) many-fermion system, 
an arbitrarily weak short-range repulsive interaction is known to 
completely destroy the conventional Fermi liquid, thereby giving rise to 
the so-called Luttinger behavior. As one of the hallmarks of the Luttinger 
regime, the electron propagator exhibits an algebraic decay
with distance which is controlled by 
a non-universal exponent and appears to be faster than in the non-interacting case.

In higher dimensions (3D and 4D) the Fermi liquid is believed to be more robust, 
although it is not expected to remain absolutely stable. In the absence of 
a clear-cut evidence, however, the possibility of a non-Fermi liquid behavior in $D>2$ 
has recently become the subject of an intense debate.

While in the case of short-ranged repulsive interactions any departures from the Fermi liquid 
are likely to be limited to the infinitely strong coupling regime, 
the long-ranged forces are considered to be capable of destroying the Fermi liquid even at finite coupling.
One of the most extensively studied such examples, the finite density system of 
non-relativistic massive fermions which are minimally coupled to an abelian gauge field,   
was indeed shown to manifest a distinctly non-Fermi liquid behavior, 
although the latter turned out to be rather different from the Luttinger one \cite{Reizer}.

More recently, there has been a renewed interest in the 3D relativistic 
counterpart of this model governed by the conventional 
$QED_3$ action 
\be
S[\psi,{\overline \psi},{A}]=
\int d{\bf z}[\sum_{f=1}^N{\overline \psi}_f
(i{\hat \gamma}^\mu \partial_\mu+{\hat \gamma}^\mu 
A_\mu-m)\psi_f+{1\over 4g^2}(\partial_\mu A_\nu - \partial_\nu A_\mu)^2]
\ee 
for the $N$-flavored Dirac fermions which are described by a reducible (four-component)
representation corresponding to the choice of the $\gamma$-matrices 
${\hat \gamma}_\mu={\hat \sigma}_\mu\otimes{\hat \sigma}_3$ 
constructed from the triplet $\sigma_\mu$ of Pauli matrices.
The use of the parity-even representation guarantees that
radiation corrections do not generate any parity-breaking Chern-Simons terms.

Among the previously discussed examples of the 3D condensed matter 
systems that support the Dirac-like low-energy excitations 
and allow for such an effective description are 
the so-called flux phase in doped Mott insulators \cite{Mavromatos,Lee}
and $d$-wave superconductors with strong phase fluctuations \cite{Wen1,Franz1,Ye1,Herbut1} 
proposed as an explanation of the pseudogap and insulating (spin and/or charge 
density wave) phases of the high-$T_c$ cuprates. Also, the non-Lorentz-invariant version
of $QED_{3}$ was shown to provide a convenient description of the normal 
semimetalic state of highly oriented pyrolytic graphite \cite{Guinea,DVK1,Miransky}.

The number $N$ of the fermion flavors depends on the problem in question,
although it is not necessarily equal to
the number of different conical Dirac points in the bare electron spectrum. 
In the abovementioned condensed matter-inspired $QED$-like models 
\cite{Mavromatos,Lee,Wen1,Franz1,Ye1,Herbut1,Guinea,DVK1,Miransky}, 
$N=2$ is a number of the electron spin components, while the 
number of the conical points turns out to be either four \cite{Wen1,Franz1,Ye1,Herbut1} 
or two \cite{Guinea,DVK1,Miransky}.

In some of the strongly correlated electron systems 
described by the effective  $QED$-like theories, such as the problem of graphite, 
the theory is formulated in terms of the electrons themselves
which enables one to directly compute various experimentally 
relevant observables \cite{DVK1,Miransky}.

However, in other systems the effective gauge fields serve as 
a convenient representation of such bosonic collective modes as spin 
or pairing fluctuations, while the Dirac fermions correspond to 
some auxiliary fermionic excitations, such as spinons \cite{Mavromatos,Lee,Wen1},
"topological" fermions \cite{Franz1,Ye1,Herbut1}, and so forth. 

In the latter case, a generic quantum mechanical 
amplitude written in terms of the Lagrangian fermions 
turns out to be gauge-dependent, while all the physical observables, which 
experimental probes can only couple to, must be manifestly gauge-invariant.   
In light of that, it is absolutely imperative to establish 
the correct form of the physical electron propagator in terms of the Lagrangian fermions.
Until this task is accomplished, no theoretical 
prediction obtained in the framework of the $QED$-like effective models
can be put under a decisive test against experimental data
which will be the ultimate means of ascertaining the validity of these scenarios.

\section{Exact versus limited gauge invariance of fermion amplitudes}

In the recent theory of the 
pseudogap phase of the high-$T_c$ superconductors \cite{Wen1,Franz1,Ye1},  
the authors advocated the use of the "stringy ansatz" with the inserted 
Wilson-like exponent of the line integral 
\be
G_\Gamma(x-y)=<0|\psi(x)\exp(-i\int_\Gamma A_\mu(z)dz^\mu){\overline \psi}(y)|0> 
\ee
as a viable candidate for the gauge-invariant physical electron propagator
(in spite of its being gauge-independent, the function
$G_{\Gamma}$ explicitly depends on the choice of the contour $\Gamma$
which connects the end points $x$ and $y$).

Moreover, the authors of Refs.\cite{Wen1,Franz1,Ye1} 
argued that the amplitude (2) exhibits the so-called Luttinger-type behavior 
\be
G_\Gamma(x)\propto {{\hat x}\over |x|^{d+\eta}}
\ee
characterized by a $positive$ (albeit varying from one reference to another)
value of the anomalous exponent $\eta$, 
provided that the contour $\Gamma$ is chosen as the straight-line segment between the end points. 

Although the idea of using Eq.(2) for elucidating the 
behavior of the physical electron propagator has been previously entertained, one must 
be cautioned by the fact that, apart from its relative simplicity,
it never received any firm justification. A potential problem 
with both the simple form of the line integral and the particular 
choice of the contour $\Gamma$ in Eq.(2) stem from 
the fact that in the theory of Refs.\cite{Wen1,Franz1,Ye1} 
the phase $\phi(x)$ of the (singular) gauge transformation 
$\Psi(x)=e^{i\phi(x)}\psi(x)$ which relates the physical electron operator $\Psi(x)$ 
to the Lagrangian fermion field $\psi(x)$ is not uniquely defined. 

In what follows, we address some of the misconceptions pertaining to 
the use of Eq.(2) in the earlier works 
and ascertain the status of the previously obtained results. 
To this end, we start out by casting $G_\Gamma(x-y)$ in the form of a functional integral
taken over both the gauge and fermion Lagrangian variables
\be
G_\Gamma(x-y)=\int D[{A}]D[{\overline \psi}]D[{\psi}] \psi(x)
{\overline \psi}(y)\exp(i\int_\Gamma A_\mu(z)dz^\mu)
\exp(iS[\psi,{\overline \psi},{A}])
\ee
Upon integrating the fermions out, the amplitude (4) turns into a functional average
over different gauge field configurations 
\be
{G}_\Gamma(x-y)= \int D[{A}]{\cal G}(x, y|{A})
\exp(i\int_\Gamma A_\mu(z)dz^\mu)\exp(iS_{eff}[{A}])
\ee
applied to the inverse Dirac operator 
${\cal G}(x,y|A)=<x|[i{\hat \partial}+{\hat A}(z)-m]^{-1}|y>$ 
computed for a given gauge field configuration $A_\mu(z)$. 

The weight of the average in Eq.(5)  
is controlled by the effective gauge field action  
\be
S_{eff}[{A}]=
{1\over 4g^2}\int d{\bf x}( \partial_\mu A_\nu - \partial_\nu A_\mu )^{2} +
{1\over 2}\int d{\bf x}\int d{\bf y} A_\mu(x)\Pi_{\mu\nu}(x-y)A_\nu(y)+\dots
\ee
where the dots stand for the higher order (non-Gaussian) terms produced by 
fermion polarization (processes of "light-light scattering" and alike) 
which we hereafter neglect, as in all of the previous works on the subject.

With the transverse fermion 
polarization $\Pi_{\mu\nu}(q)=\Pi(q)(\delta_{\mu\nu}-q_\mu q_\nu/q^2)$ taken in to account,
the gauge field propagator computed in the covariant $\lambda$-gauge assumes the form 
\be
{D}_{\mu\nu}(q)={g^2\over {q^2+g^2\Pi(q)}}(\delta_{\mu\nu}+(\lambda-1){q_\mu q_\nu\over q^2})
\ee
In order to obtain Eq.(7) one has to complement (6) with a (generally, non-local) gauge fixing term
$\Delta S_{eff}[A]=\int d{\bf x}\int d{\bf y}\partial_\mu A_\mu(x)
<x|[1-g^2\Pi(i\partial)/\partial^2]|y>\partial_\nu A_\nu(y)/2g^2\lambda$.

In the case of the conventional (that is weakly coupled, $g^2\ll 1$) $QED_4$,
the fermion polarization behaves as $\Pi(q)\propto q^2$, and the above gauge
fixing term becomes local in the coordinate space.
As a result, the ultraviolet (UV) divergencies 
occur at fermion momenta of order the upper momentum cutoff. 

In contrast, $QED_3$ develops its highly non-trivial 
behavior at $\it intermediate$ fermion momenta
$m\ll p\ll\Lambda=Ng^2$ controlled by the (this time, dimensionful) coupling constant $g$
\cite{Mavromatos}.
However, adhering to the customary terminology, we will refer to this regime 
as the UV one, for this is where all
the logarithmic corrections originate from, whereas 
above the cutoff $\Lambda$ no terms $\propto\ln(\Lambda/p)$ can possibly occur.

In this regime, the gauge propagator (7) 
is totally dominated by the fermion polarization which, 
due to the parity conserving structure of the reducible four-fermion representation
produces no Chern-Simons terms and, in the leading $1/N$ approximation, reads as  
\be
\Pi(q)={N\over 8}{\sqrt {q^2}}
\ee
In this way, the number of fermion flavors $N$ becomes the actual parameter 
controlling perturbation expansion, regardless of the strength of the bare coupling $g$. 

Without making any additional approximations, 
the function ${\cal G}(x,y|A)$ can be expressed in the form of a quantum mechanical 
path integral \cite{Stefanis} (see also \cite{DVK3} where this representation was used in
a non-perturbative calculation of the Dirac fermion propagator 
in a $\it static$ spatially random gauge field which is pertinent to such problems 
as the effect of vortex disorder on the quasiparticle properties of $d$-wave superconductors 
or that of dislocations in layered graphite)
\be
{\cal G}(x,y|{A})=\int^\infty_0 d\tau\int_{x(0)=y}^{x(\tau)=x}
D{x}D{p}e^{i{\hat S}_0[x,p]}\exp(i\int^\tau_0d\tau^\prime{A}_\mu(z){dz^\mu\over d\tau^\prime})
\ee
where $x_\mu(\tau)$ is a fermion trajectory parametrized by the proper time
$\tau$. The spinor structure of the fermion propagator is fully accounted for 
by integrating over the additional variable $p_\mu(\tau)$ with the free 
(matrix-valued) fermion action 
\be
{\hat S}_0[x(\tau), p(\tau)]=\int^\tau_0 d\tau^\prime 
{p}_\mu[{d{x}^\mu\over d\tau^\prime}-{\hat \gamma}^\mu -m],
\ee
thus providing a systematic improvement of the celebrated Bloch-Nordsieck model. In the latter, 
all the spin-related effects are ignored which makes
this model exactly soluble but restricts its applicability to
the infrared (IR) regime $|p^2-m^2|\ll m^2$ near the fermion mass shell.
In the IR regime, the relevant fermion trajectories 
contributing to (9) deviate only slightly from the straight-path contour $\Gamma$
(which coincides with the fermion world line in the case 
of a light cone-like separation between the end points, $(x-y)^2=0$).

One must recognize that the IR regime can only exist if the fermions are massive,
while in the case of $m=0$ the entire region below the upper cutoff $\Lambda$ 
falls into the opposite, UV, regime.
In the UV regime, the fermion trajectories which dominate the amplitude
(9) may strongly depart from the straight-path contour $\Gamma$,
for there is no mass term to suppress such deviations.

Combining Eqs.(5) and (9) we represent the amplitude (2) in the form 
manifesting its gauge invariance
\be
G_\Gamma(x-y)=\int^\infty_0 d\tau\int_{x(0)=y}^{x(\tau)=x}
Dx Dp e^{i{\hat S}_0[x,p]}<\exp(i\oint\partial_\mu A_\nu d\Sigma^{\mu\nu})>
\ee
where the brackets stand for the (normalized) functional average 
over the gauge field with the weight $\exp(iS_{eff}[A])$ which is to be 
taken for each trajectory $x_\mu(\tau)$ before the sum over all the 
different trajectories is executed.

In order to obtain the area integral in the 
exponent we applied Stokes' theorem to the line integral
taken along a closed contour which is composed of a trajectory $x_\mu(\tau)$ 
and the straight-line segment $\Gamma$, parametrized 
as $x^{(0)}_\mu(\tau^\prime)=\tau^\prime (x-y)_\mu/\tau$ and traced backwards. 

The averaging procedure generates a sum of all the multi-loop diagrams with no 
couplings between the fermion polarization insertions into the gauge field propagators
and the open fermion line which describes the fermion propagating from $y$ to $x$.

Elucidating the structure of the gauge invariant amplitude (2) 
helps one to appreciate the difference between (11) and all the other ($\xi\neq 0$)  
members of the family of functions
labeled by a continuous parameter $\xi$ (not to be confused with the covariant gauge 
parameter $\lambda$!) 
$$
G_\xi(x-y)={<0|\psi(x)\exp(i(\xi-1)\int_\Gamma A_\mu(z)dz^\mu)
{\overline \psi}(y)|0>\over <0|\exp(i\xi\int_\Gamma A_\mu(z)dz^\mu)|0>}=
$$
\be
\int^\infty_0d\tau\int_{x(0)=y}^{x(\tau)=x}DxDp e^{iS_0[x,p]}
{<\exp(i\xi\int_\Gamma A_\mu dz^\mu +i\oint \partial_\mu A_\nu d\Sigma^{\mu\nu})>
\over <\exp(i\xi\int_\Gamma A_\mu dz^\mu)>}
\ee 
Amongst others, Eq.(12) also includes the function 
\be
G_1(x-y)={<0|\psi(x){\overline \psi}(y)|0>\over <0|\exp(i\int_\Gamma A_\mu(z)dz^\mu)|0>}
\ee
which was invoked by the authors of Ref.\cite{Franz1}
who asserted that $G_1(x)$ given by Eq.(13) is identical to the amplitude (3)
and, therefore, can be used to compute $G_\Gamma(x)=G_0(x)$.

The UV anomalous dimension of Eq.(13) can be deduced rather straightforwardly 
by computing the ratio between the wave function renormalization factor 
determining the anomalous dimension of the ordinary (gauge variant) 
fermion propagator \cite{Nash}
\be
<0|\psi(x){\overline \psi}(y)|0>\propto G(x-y)(\Lambda |x-y|)^{(4/3\pi^2N)(2-3\lambda)}
\ee
(hereafter $G(x)$ stands for the free fermion propagator which 
also yields the bare value of $G_\xi(x)$ for any $\xi$) 
and the Gaussian average of the Wilson line
\be
<\exp(i\int_\Gamma A_\mu dz^\mu)>=
\exp[-{1\over 2}\int_\Gamma dz_1^\mu \int_\Gamma dz_2^\nu D_{\mu\nu}(z_1-z_2)]\propto
(\Lambda |x-y|)^{(4/\pi^2N)(2-\lambda)},
\ee
thus indeed resulting in the anticipated Luttinger-type behavior of Eq.(13) 
which is characterized by the overall $positive$ anomalous dimension 
\be
\eta^{3d}_1={16\over 3\pi^2N}
\ee
Notably, this result is free of the gauge parameter $\lambda$, thus creating the impression
that Eq.(13) represents a truly gauge invariant function. 

As one immediate objection to the assertion of Ref.\cite{Franz1},
one might recall that the very nature of the relationship between the electrons 
and the auxiliary Lagrangian fermions implemented via the singular
gauge transformation  
$
<0|\Psi(x){\overline \Psi}(y)|0>=<0|\psi(x)\exp[i(\phi(x)-\phi(y)]{\overline \psi}(y)|0>
$
implies that the physical electron propagator 
can only be given by a $\it single$ average over the gauge field, rather than
a $\it ratio$ of such. 

However, except for the case of the original amplitude $G_0(x)$,
Eq.(12) is represented by a ratio of the two averages and, therefore, 
it can not be truly gauge invariant.
Indeed, the gauge fields, over which one averages in both the 
numerator and denominator of Eq.(12), may transform totally independently of one another 
($A^{1,2}_\mu\to A^{1,2}_\mu+\partial_\mu f^{1,2}$),
thus resulting in the overall phase factor $\exp[i\xi(f^1(x)-f^2(x)-f^1(y)+f^2(y))]$ 
which only vanishes for $f^1(x)=f^2(x)$, as long as $\xi\neq 0$. 

Nevertheless, as shown below, $G_\xi(x)$ turns out to be independent of 
the gauge parameter $\lambda$ within the class of the covariant gauges for any $\xi$.
Presumably, it was this $limited$ gauge independence that led the 
authors of Ref.\cite{Franz1} to their conclusion that  
the (seemingly gauge-invariant) amplitude
$G_1(x)$ must be identical to the (truly invariant) amplitude $G_0(x)$, for they apparently
coincide with each other in the axial gauge $(x-y)^\mu A_\mu(z)=0$, 
in which the line integral $\int_\Gamma A_\mu(z) dz^\mu$ vanishes. 

However, should this happen to be valid, the same argument
would also apply to an arbitrary $G_\xi(x)$
which is seemingly gauge invariant to absolutely the same extent as 
Eq.(13) is (that is, provided that one uses the $identical$ gauge transformation in 
both numerator and denominator), and all these functions
coincide with each other when computed in the axial gauge. 

Nonetheless, one can readily see that $G_\xi(x)$ does bear a non-trivial $\xi$-dependence
by simply evaluating the integrand in Eq.(12) for certain chosen trajectories. 
As an example, we consider the case of a square-shaped
closed contour consisting of four straight segments, each of length $|x-y|$. Having 
all the adjoint sides of the square at right angles to each other guarantees one
from a possible occurrence of the additional ("cusp") UV singularities \cite{Stefanis}.

Applying the method of dimensional regularization near $d=3$ we obtain 
$$
{<\exp(i\oint \partial_\mu A_\nu d\Sigma_{\mu\nu}+
i\xi\int_\Gamma A_\mu dz^\mu)>\over <\exp(i\xi\int_\Gamma A_\mu dz^\mu>}\propto 
$$
$$
\exp[-{4(\Lambda|x-y|)^{3-d}\over \pi^{(d+1)/2}N}\Gamma({d-1\over 2})(2-\xi)
[I_1(1-(\lambda-1)(d-2))-2I_2(\lambda-1)(d-1)]]$$
\be
\propto(\Lambda |x-y|)^{8(2-\xi)/\pi^2N}
\ee
where the terms in the exponent proportional to  
\be
I_1=\int^1_0\int^1_0 {d\alpha d\beta\over 
|\alpha-\beta|^{d-1}}={2\over (3-d)(2-d)}
\ee
and 
\be
I_2=\int^1_0\int^1_0{\alpha\beta d\alpha d\beta\over 
(\alpha^2+\beta^2)^{(1+d)/2}}={2\over (3-d)(1-d)}[2^{(1-d)/2}-1]
\ee
correspond to the line integrals
taken along the same and two adjoint sides of the square, respectively
(pairs of the opposite sides produce no terms $\propto\ln (\Lambda |x-y|)$ 
which is what we are after). 
Despite the fact that the gauge parameter $\lambda$ cancels out,
as expected, the values of the exponent given by Eq.(17)
are markedly different for, say, $\xi=0$ and $\xi=1$.

In fact, Eq.(17) can either vanish or grow with increasing $|x-y|$,
depending merely on whether $\xi$ is greater or smaller than $2$.
This observation suggests that for different values of $\xi$ the functions $G_\xi(x)$  
are indeed different (it would be rather 
fortuitous if the $\it sum$ over all the fermion trajectories were $\xi$-independent, 
despite the difference between contributions from each $individual$ trajectory).

It is worth mentioning that in the case of massive fermions
the different functions (12) do become approximately identical
in the vicinity of the mass shell ($|p^2-m^2|\ll m^2$)
where the leading functional dependence on the gauge  
field originates from the eikonal phase factor ${\cal G}(x,y|A)\propto
\exp(i\int_\Gamma A_\mu dz^\mu)$. Therefore,  
the entire IR divergence of the ordinary (gauge variant) 
propagator can be obtained by simply averaging this exponential factor, 
while any $G_\xi(x)$ appears to be totally IR divergence-free.

This factorization of the exponent of the line integral in the IR regime 
which had long been known in the case of $QED_4$
\cite{Brown} was extended into the $3D$ case in Ref.\cite{DVK1}, where it was shown  
that for any $\xi$ the function (12) possesses the same simple pole
in the momentum representation ${G}_\xi(p)\propto({\hat p}-m)/(p^2-m^2)$, 
which corresponds to the universal ($\xi$-independent) behavior in the coordinate space 
$G_\xi(x)\propto e^{-m|x|}$. 
 
In order to avoid confusion we reiterate that the universal IR behavior by
no means prevents the different functions (12) from having  
different UV anomalous dimensions, as manifested by Eq.(17).  
Incidentally, in the massless case
it turns out to be the UV exponent $\eta_\xi$ which controls the 
power-law decay (3) extending to arbitrarily long distances. 

\section{Perturbative calculation of fermion amplitudes}

Having demonstrated the possibility of a non-trivial $\xi$-dependence non-perturbatively, 
we now present a direct perturbative calculation of $G_\xi(x)$. 
Expanding (12) to second order in $A_\mu(z)$ 
we find three different kinds of correction terms
\be
G_\xi(x-y)=G(x-y)-\int d{\bf z}_1 d{\bf z}_2
<G(x-z_1){\hat A}(z_1)G(z_1-z_2){\hat A}(z_2)G(z_2-y)>+
\ee
$$
+(\xi-{1\over 2})G(x-y)<\int_\Gamma A_\mu(z_1) dz_1^\mu\int_\Gamma A_\nu(z_2)dz_2^\mu>+
(1-\xi)\int d{\bf z}_1<G(x-z_1){\hat A}(z_1)G(z_1-y)\int_\Gamma A_\mu(z_2) dz_2^\mu>
$$
The first term in the r.h.s. of (20) corresponds to
the first order self-energy correction to the ordinary propagator which 
is given by the leading term of $1/N$ expansion of Eq.(14) 
\be
\delta_1G_\xi(x-y)=-\int d{\bf z}_1 d{\bf z}_2 G(x-z_1)\gamma^\mu 
D_{\mu\nu}(z_1-z_2)\gamma^\nu G(z_2-y)=
G(x-y){4\over 3\pi^2N}(2-3\lambda)\ln(\Lambda |x-y|)
\ee
The second term in (20) originates from $1/N$ expansion of the Wilson line (15) 
inserted into the numerator and denominator of Eq.(12) 
with the prefactors $(\xi-1)$ and $\xi$, respectively
\be
\delta_2G_\xi(x-y)=(\xi-{1\over 2})G(x-y)\int_\Gamma dz_1^\mu \int_\Gamma dz_2^\nu 
D_{\mu\nu}(z_1-z_2)=G(x-y){4\over \pi^2N}(1-2\xi)(2-\lambda)\ln(\Lambda |x-y|)
\ee
The third term in (20) stems from expanding both $G(x,y|A)$ and the Wilson line
in the numerator of Eq.(12) to first order in $A_\mu(z)$ 
$$
\delta_3G_\xi(x-y)=(1-\xi)\int d{\bf z}_1G(x-z_1)\gamma^\mu G(z_1-y)\int_\Gamma dz_2^\nu
D_{\mu\nu}(z_1-z_2)\approx
$$ 
$$
\approx(1-\xi)\int d{\bf z}_1[G(x-y)\gamma^\mu G(z_1-y)+G(x-z_1)
\gamma^\mu G(x-y)]\int_\Gamma dz_2^\nu
D_{\mu\nu}(z_1-z_2)=
$$
$$
={2\over \pi^3N}(1-\xi)G(x-y)\gamma^\mu\int_\Gamma dz_2^\nu\int d{\bf z}_1
{({\hat z}_1-{\hat y})\over |z_1-y|^3}{1\over |z_1-z_2|^2}
[\lambda\delta_{\mu\nu}+2(1-\lambda){(z_1^\mu-z_2^\mu)(z_1^\nu-z_2^\nu)\over
|z_1-z_2|^2}]
=
$$
\be
=G(x-y){8\over \pi^2N}(1-\xi)\lambda\ln(\Lambda |x-y|)
\ee
This result was obtained with logarithmic accuracy, and 
in the course of the calculation we used the coordinate space representation 
of the 3D propagators
\be
G(x)={1\over 4\pi}{{\hat x}\over x^3}, ~~~~~D_{\mu\nu}(x)={4\over \pi^2N|x|^2}
[\lambda\delta_{\mu\nu}+2(1-\lambda){x^\mu x^\nu\over |x|^2}]
\ee
and the following $d$-dimensional integrals 
\be
\int d{\bf x}{x^\alpha\over |x|^d|x-y|^{d-1}}={2\pi^{d/2}\over \Gamma(d/2)}
{y^\alpha\over |y|^{d-1}} 
\ee
and
\be
\int d{\bf x}{x^\alpha(x^\beta-y^\beta)(x^\gamma-y^\gamma)
\over |x|^d|x-y|^{d+1}}={2\pi^{d/2}\over 3(d-1)\Gamma(d/2)}
[{2y^\alpha\delta^{\beta\gamma}-y^\beta\delta^{\alpha\gamma}-y^\gamma\delta^{\alpha\beta}
\over |y|^{d-1}} 
+
(d-1){y^\alpha y^\beta y^\gamma\over |y|^{d+1}}]
\ee
The logarithmic dependence in Eqs.(21-23) results from taking the 
$d\to 3$ limit of the line integral
$\int_\Gamma dz^\mu(z-y)^\mu/|z-y|^{d-1}=(\Lambda |x-y|)^{3-d}/(3-d)\to\ln(\Lambda|x-y|)$.

Combining Eqs.(21-23) together we read off the overall UV anomalous dimension  
\be
\eta^{3d}_\xi={16\over 3\pi^2N}(3\xi-2)
\ee
In particular, for $\xi=1$ we reproduce Eq.(16), while for $\xi=0$ one obtains
\be
\eta^{3d}_0=-{32\over 3\pi^2N}
\ee
Thus, despite the fact that the gauge parameter $\lambda$ cancels out, as expected,
the functions $G_\xi(x)$ are indeed different for different values of 
the parameter $\xi$. 

At first sight, the non-trivial dependence 
of Eq.(27) on the parameter $\xi$ (which is linear to first order in $1/N$) 
may seem to be in conflict with the fact that all the functions $G_\xi(x)$
coincide with each other when computed in the axial gauge.

In fact, there would only be a contradiction, if Eq.(12) were truly gauge invariant.
Instead, albeit invariant within the class of covariant gauges,
the function $G_\xi(x)$ may take a different value in the axial gauge,
for the latter does not belong to this class, for any $\xi\neq 0$.

On the other hand, 
the unique property of the function $G_0(x-y)$ which is given by a $single$ average
instead of a $ratio$ of the two does guarantee its true gauge invariance
and the possibility to obtain Eq.(28) in arbitrary gauge, including both 
the covariant and the axial ones. 
In fact, it was shown in Ref.\cite{DVK2} that these two calculations do agree with each other,
and yet another independent calculation employing
the so-called "radial" (Fock-Schwinger) gauge $(x-y)^\mu A_\mu(x)=0$ \cite{Gusynin1} 
provides further support for this conclusion.

However, the finding that the UV exponent $\eta^{3d}_0$ assumes the negative value 
suffices to disqualify the "stringy ansatz" (2) from being
a sound candidate for the physical electron propagator, since,  
in all of the $QED$-like models discussed so far,
repulsive electron-electron interactions are expected
to result in a suppression, rather than enhancement, of any amplitude
describing propagation of physical electrons.

The above analysis can be easily extended to the case of the conventional
weak coupling $QED_4$ which demonstrate that the situation in 3D is not at all exceptional.
Instead of Eqs.(21-23) we now get 
\be
\delta_1G_\xi(x-y)=-G(x-y){g^2\over 8\pi^2}\lambda\ln(\Lambda |x-y|)
\ee
\be
\delta_2G_\xi(x-y)=G(x-y){g^2\over 8\pi^2}(1-2\xi)(3-\lambda)\ln(\Lambda |x-y|)
\ee
\be
\delta_3G_\xi(x-y)=G(x-y){g^2\over 4\pi^2}(1-\xi)\lambda \ln(\Lambda |x-y|)
\ee
and, instead of Eqs.(24-26) , we use the formulas
\be
G(x)={1\over 2\pi^2}{{\hat x}\over x^4}, ~~~~~D_{\mu\nu}(x)={g^2\over 4\pi^2|x|^2}
[{1+\lambda\over 2}\delta_{\mu\nu}+(1-\lambda){x^\mu x^\nu\over |x|^2}]
\ee
\be
\int d{\bf x}{x^\alpha\over |x|^d|x-y|^{d-2}}={\pi^{d/2}\over \Gamma(d/2)}
{y^\alpha\over |y|^{d-2}} 
\ee
\be
\int d{\bf x}{x^\alpha(x^\beta-y^\beta)(x^\gamma-y^\gamma)
\over |x|^d|x-y|^{d}}={\pi^{d/2}\over 2(d-2)\Gamma(d/2)}
[{y^\alpha\delta^{\beta\gamma}-y^\beta\delta^{\alpha\gamma}-y^\gamma\delta^{\alpha\beta}
\over |y|^{d-2}} 
+
(d-2){y^\alpha y^\beta y^\gamma\over |y|^{d}}]
\ee
Combining Eqs.(29-31) together we obtain the UV anomalous dimension
of $G_\xi(x)$ in the 4D case 
\be
\eta^{4d}_\xi={3g^2\over 8\pi^2}(2\xi-1)
\ee 
Interestingly enough, for $\xi=0$ and $\xi=1$ the values of Eq.(35) differ only in their sign,
and the anomalous dimension $\eta_0^{4d}=-3g^2/8\pi^2$ is again negative.

\section{Alternate forms of physical electron propagator}

The apparent problem with the unphysical 
decay of the conjectured form of the physical electron propagator
(2) which appears to be slower than in the case of non-interacting electrons 
compels one to explore alternate proposals. To this end, one may be able to benefit
from the previous studies of the conventional $QED_4$ where the problem 
of constructing gauge invariant asymptotical quantum states 
with quantum numbers of an electron
("dressed charges") was long recognized, and attempts to explicitly solve  
it have long been under way. 

A (non-local) composite operator creating such a state can be sought in the form 
\be
\Psi(x)=\exp[i\int d{\bf y}f^\mu(x-y)A_\mu(y)]\psi(x)
\ee
where the vector function $f^\mu(x)$ is subject to the condition 
$\partial_\mu f^\mu(x)=\delta(x)$ imposed by the requirement of gauge invariance.

In the original proposal which was put forward by 
Dirac in his pioneering work of the fifties, the physical electron was identified
with the operator 
\be
\Psi_D(x)=\exp(i{\nabla_iA_i\over \nabla^2})\psi(x)
\ee
Indeed, the expectation value $<0|\Psi_D(x){A_\mu}(y)\Psi_D^\dagger(x)|0>=
\delta_{\mu 0}/4\pi|{\vec x}-{\vec y}|$ describes the Coulombic 
electrostatic field, thus suggesting that Eq.(37) provides a plausible 
representation of a $\it static$ charge.

Moreover, it was found that the Fourier transform of the gauge-invariant amplitude 
$<0|\Psi_D(x){\overline \Psi}_D(y)|0>$ is IR finite at the single point 
$p_\mu=(m, {\vec 0})$ on the mass shell \cite{Lavelle}. 

It was also argued in Ref.\cite{Lavelle} that, once created, the coherent state 
$\Psi^\dagger_D(x)|0>$ remains stable, unlike that produced by the operator
$\Psi_T(x)=\exp(i\int^t_{-\infty}A_0(t^\prime)dt^\prime)\psi(x)$ whose propagator
$<0|\Psi_T(x){\overline \Psi}_T(y)|0>$ is given by Eq.(2) where
$(x-y)_\mu\propto\delta_{\mu 0}$.

Furthermore, the amplitude $<0|\Psi_D(x){\overline \Psi}_D(y)|0>$ was found to undergo  
multiplicative UV renormalization, featuring a $\it positive$ UV exponent
\be
\eta^{4d}_D=g^2/8\pi^2
\ee
The authors of Ref.\cite{Lavelle}
were also able to generalize the construction (37) onto the case of a moving charge 
\be
\Psi_v(x)=\exp(i{\delta_{\mu\nu}-({u}+v)_\mu({u}-v)_\nu
\over 
\partial_\mu^2-({u}_\mu\partial_\mu)^2+(v_\nu\partial_\nu)^2}\partial_\mu A_\nu)\psi(x)
\ee
where ${u}_\mu=(1,{\vec 0})$ is the time-like unit vector and $v_\mu=(0,{\vec v})$
is the velocity of the charge. 
The formula (39) was obtained by requiring
that, in accord with the anticipated physical interpretation of Eq.(39),
the electromagnetic field $<0|\Psi_v(x){A_\mu}(y)\Psi_v^\dagger(x)|0>$ 
associated with the coherent state created by this composite operator 
reproduces the classical Lienard-Wiechert potentials.

It was also found that the corresponding propagator
\be
G_v(x-y)=<0|\psi(x)\exp[i{\delta^{\mu\nu}-({u}^\mu+v^\mu)({u}^\nu-v^\nu)\over 
\partial_\alpha^2-({u}_\alpha\partial_\alpha)^2+
(v_\alpha\partial_\alpha)^2}\partial_\mu A_\nu]
{\overline \psi}(y)|0>
\ee
is both IR-finite at $p_\mu =m(1, {\vec v})/{\sqrt {1-v^2}}$
and multiplicatively UV-renormalizable, its anomalous dimension being explicitly $v$-dependent
\cite{Lavelle}
\be
\eta^{4d}_v=-{g^2\over 8\pi^2}(3+2{1\over v}\ln({1-v\over 1+v}))
\ee 
The combination of all the befomentioned 
properties makes Eq.(39) the best (up-to-date) available candidate 
for the physical electron operator in the conventional ($\it massive$) $QED_4$. 

It is worth mentioning that 
Eq.(41) coincides with the UV dimension of the $4D$ "stringy" amplitude (2)
only in the unphysical limit $v\to\infty$, while its $v=0$ value yields Eq.(38).

By construction, Eq.(41) can not be applied directly to the 
$\it massless$, let alone the $3D$ massless, case which corresponds to the limit $v\to 1$
where the UV dimension (41) increases towards infinitely high $\it positive$ values.

Albeit not being immediately applicable to the case of $m=0$, the calculation 
of $G_v(x)$ carried out in Ref.\cite{Lavelle} suggests that 
the logarithmic growth of (41) gets effectively cut off at $max(|1-v|, 1/\Lambda|x|)$.
Therefore, it is not inconceivable
that the massless counterpart of Eq.(40) may exhibit a faster than a power-law decay
\be
G_{phys}(x)\propto\exp(-const\ln^2(\Lambda|x|)),
\ee
where the constant is proportional to either $1/N$ (3D) or $g^2$ (4D),
thus placing the effective $QED$-like theories 
into the class of "super-Luttinger" models,
alongside the one-dimensional metals with unscreened Coulombic interactions where 
$G_{phys}(x)\propto\exp(-const\ln^{3/2}(\Lambda |x|))$ \cite{Schulz}.

In light of this possibility, the previous claims of discovering  
the Luttinger-like behavior in the framework of the $QED_3$-theory of the
pseudogap phase \cite{Wen1,Franz1,Ye1} 
may need to be prepared to handle a potentially much stronger suppression 
of the physical amplitudes in order to reconcile such a behavior  
with the available photoemission, tunneling, and other experimental data.

\section{Discussion}

Evidently, one could escape the whole problem
with the singular massless limit, if the Dirac fermions acquired a 
finite mass via the mechanism of chiral symmetry breaking \cite{Appelquist1}. 
In the context of the $QED_3$ theory 
of the pseudogap phase, such an intrinsic instability is expected to result in the onset of
a spin (SDW) and/or charge (CDW) density wave  
\cite{Herbut1,Franz3}, provided that such a transition
does occur for the physical number of fermion flavors ($N=2$). 

Thus far, there have been only preliminary conclusions drawn on the role of the 
strong spatial anisotropy of the quasiparticle dispersion 
($\epsilon={\sqrt {v_1^2q_1^2+v_2^2q_2^2}}$ with $v_1/v_2>10$) 
in the high-$T_c$ cuprates.

Although, in accord with general physical expectations, 
this anisotropy was found to decrease upon IR renormalization,\cite{Franz3,Herbut2},
there has been no consensus reached, regarding universality (or a lack thereof) 
of the critical number of flavors $N_{cr}$, 
much less regarding the possibility that the spatial anisotropy can 
drive this critical number to the values in excess of $2$. 

To this end, it may be helpful 
to complement the renormalization group approach adopted in Refs.\cite{Franz3,Herbut2} by 
a direct analysis of the strong coupling regime
where the gauge field propagator
(7) is dominated by the fermion polarization $\Pi(q)$. In the case of anisotropic
fermion dispersion, the latter becomes a function of the 
spatial components of the momentum $\vec q$ multiplied by the corresponding velocities
$v_iq_i$, thus allowing one to scale such factors out in the Schwinger-Dyson
equation for the fermion mass \cite{Appelquist1} and suggesting that $N_{cr}$
may even remain the same as in the isotropic case.

In any event, the effect of the spatial
anisotropy of the fermion dispersion may be rather different from that 
of the Lorentz non-invariant (e.g., pure Coulombic)
couplings which can 
cause $N_{cr}$ to $\it decrease$ \cite{DVK1,Miransky}, as compared to the value
obtained in the Lorentz-invarint case.  

One should, however, be alerted by the fact that even in the perfectly Lorentz-invariant case
the status of the analyses based on the Schwinger-Dyson
equation (which, in this particular case, suggests $N_{cr}>3$) remains 
somewhat unclear.
In this regard, it was pointed out in Ref.\cite{Appelquist2} that 
the Schwinger-Dyson equation may systematically overestimate 
the actual critical number of flavors which might, in fact, be as low as $3/2$.

It is, therefore, quite remarkable
that, as an alternate route, the intrinsic SDW instability
of the pseudogap phase of the cuprates can instead     
manifest itself through a divergent staggered spin susceptibility \cite{Wen2,Gusynin2}
\be
\chi_{Q}(q)\propto ({\sqrt {q^2}})^{1-64/(3\pi^2N)}
\ee
where the momentum $\vec q$ corresponds to the deviation from 
the SDW ordering vector ${\vec Q}=(\pi,\pi)$.

In order for this behavior to occur,
the presence of $\it massless$ neutral fermions is rather necessary than problematic. 
Curiously enough, the susceptibility (43) 
displays the $\it negative$ anomalous dimension which is exactly twice the value (28).

Although this observation does not necessarily imply that the 
staggered spin susceptibility given by the average 
${\chi}_{Q}(x-y)=<{\cal G}(x,y|A){\cal G}(y,x|A)>$ 
is directly related to the product of the two averages 
$|G_0(x-y)|^2\propto<{\cal G}(x,y|A)\exp(-i\int_\Gamma A_\nu dz^\nu)>
<{\cal G}(y,x|A)\exp(i\int_\Gamma A_\mu dz^\mu)>$,  
it may help one to elucidate the real physical meaning of the 
negative anomalous dimensions exhibited by the gauge invariant amplitude (2).

It is possible that, albeit being unfit for the role of the physical electron propagator,
the amplitude (2) may still bear some important information about the
vertex functions which also determine the behavior of the gauge-invariant susceptibilities.

As far as the experimental status of the $QED_3$ theory of the cuprates
is concerned, it is presently unclear whether it can accomodate the 
phenomenon of time reversal 
symmetry breaking, as suggested by the recent experiment \cite{Kaminski}, without generating
the Chern-Simons term in the fermion polarization. 

In contrast, broken time reversal symmetry seems to be an intrinsic feature
of some alternate approaches to the pseudogap phase which, instead of the  
phase fluctuations of the local parent $d_{x^2-y^2}$-wave order parameter
\cite{Wen1,Franz1,Ye1,Herbut1}, focus on other bosonic
collective modes, such as emergent $d$-symmetrical CDW \cite{Chakravarty} or 
incipient secondary pairing ($d_{x^2-y^2}\to d_{x^2-y^2}+id_{xy}$) \cite{Sachdev,Jens}.

To this end, it is worth mentioning that the recent experimental 
reports of a genuine quantum-critical behavior in the $Ca$-doped
cuprates \cite{Deutcher} are in quantitative agreement with the predictions \cite{Jens}
based on the theory of Ref.\cite{Sachdev}.
Notably, in this alternate (Higgs-Yukawa-type) 
theory of the cuprates the anomalous dimension of the electron propagator was
found to be positive \cite{Jens,Reenders}.

We conclude by stressing that, regardless of
the status of the $QED_3$ theory of the cuprates itself, 
the problem of constructing the gauge 
invariant fermion propagator in the effective massless $QED$-like theories, thus far, 
has received a lesser attention than it, arguably, deserves.

This problem needs to be settled before one can start drawing solid,
rather than wishful, conclusions about the true behavior in these as
well as other gauge field models, including non-abelian and discrete symmetry 
(say, $Z_n$) ones.

In the present paper, we undertook an attempt to clarify this issue 
which may have already resulted in a widespread confusion, in particular, as far as the $QED_3$ theory 
of the pseudogap phase of the cuprates is concerned.
We demonstrated that the previously proposed ansatz ehxibits a clearly unphysical behavior
and, therefore, needs to be modified. In the course of this analysis we established an interesting 
property of "partial gauge invariance" of the family of functions (12) with $\xi\neq0$ 
which, we believe, might have been partly responsible for the 
erroneous conclusions drawn in \cite{Franz1}. We also conjectured an alternate form of the physical electron propagator
which decays with distance faster than any power-law, thus challenging the interpretation of the experimental data 
based on the idea of the Luttinger-like behavior in the cuprates proposed in Refs.\cite{Wen1,Franz1,Ye1}.

\section{Acknowledgements}
The author acknowledges valuable communications with V. P. Gusynin. 
This research was supported by the NSF under Grant No. DMR-0071362.

\end{document}